Density functional theory studies of MTSL nitroxide side chain conformations attached to an activation loop

Maria Grazia Concilio,[a1*] Alistair J. Fielding,[a2*] Richard Bayliss,[b] and Selena G. Burgess[b]

[a]*The Photon Science Institute and School of Chemistry, EPSRC National EPR Facility and Service, University of Manchester, Manchester, M13 9PY, United Kingdom.*
[b]*Astbury Centre for Structural and Molecular Biology, Faculty of Biological Sciences, University of Leeds, Leeds LS2 9JT, United Kingdom.*

A quantum-mechanical (QM) method rooted on density functional theory (DFT) has been employed to determine conformations of the methane-thiosulfonate spin label (MTSL) attached to a fragment extracted from the activation loop of Aurora-A kinase. The features of the calculated energy surface revealed low energy barriers between isoenergetic minima and the system could be described in a population of 76 rotamers that can be also considered for other systems since it was found that the $\chi_3$, $\chi_4$ and $\chi_5$ do not depend on the previous two dihedral angles. Conformational states obtained were seen to comparable to those obtained in the α-helix systems studied previously, indicating that the protein backbone does not affect the torsional profiles significantly and suggesting the possibility to use determined conformations for other protein systems for further modelling studies.

[1]* First corresponding author details: email mariagrazia.concilio@postgrad.manchester.ac.uk, Telephone +44(0)7769765464.
[2]* Second corresponding author details: email alistair.fielding@manchester.ac.uk, Telephone +44 (0)161-275-4660, Fax +44 (0)161-275-4598.

## 1. Introduction

The most widely employed spin label for studies of the structure and dynamics of biomolecules[1-2] through electron paramagnetic resonance (EPR) spectroscopy is the methane-thiosulfonate spin label (MTSL). Fig. 1 shows a model of the MTSL attached to fragment extracted from the activation loop of Aurora-A kinase, a Serine/Threonine protein kinase that regulates many cellular pathways and is overexpressed in a number of cancers.[3-4]

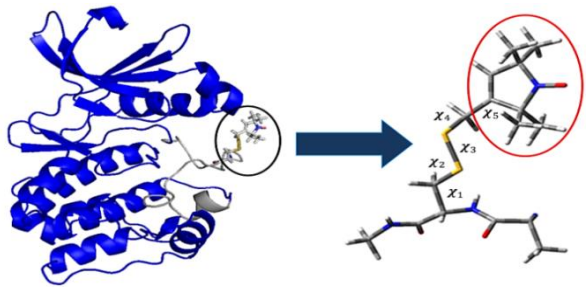

**Fig. 1:** Structure of the Aurora-A kinase domain with the MTSL side chain (black circle) attached at position 288 within the activation loop which comprises residues 274 to 299 (grey). The unit peptide extracted for the DFT analysis is indicated by a blue arrow. All five dihedral angles of the MTSL side chain are enlarged and shown, with the pyrroline ring containing the nitroxide group highlighted in a red circle.

The MTSL is endowed of high flexibility and the knowledge of the potential surface is required in order to model its internal dynamics.[5-7] In this work a quantum mechanical (QM) method based on density functional theory (DFT) was employed to determine features of energy surface of the five dihedral angles of the MTSL. The DFT theory was employed since it includes electron correlations effect, representing an appropriate method to reproduce short electrostatic interactions between sulphur atoms and backbone atoms.[7]

In the previous literature, similar approaches were adopted by Tombolato *at al.*[5] using the Hartree-Fock (HF) theory to study conformations of the MTSL in α-helix systems of the T4 Lysozyme protein[5-7] and results obtained were used to complement subsequent MD studies.[8-9] Considering our system, the question arose whether the conformational states determined in the α-helix are the same in the activation loop of a different protein.

We performed a conformational analysis of the potential energy surfaces of the MTSL side chain attached to a fragment extracted from activation loop of Aurora-A kinase protein with the purpose to characterize its geometrical parameters and describe the system in a limited number of rotamers. We carried out this work to establish a basis for more advanced modelling approaches involving molecular dynamics (MD) simulations that require different initial starting conformations of the MTSL for appropriate statistical analyses.

## 1. Methods

### 1.1 QM calculations of the conformations of the MTSL

A short unit peptide was extracted from the X-ray crystal structure of the Aurora-A kinase domain (residues 122-403 C290A C393A; PDB 4CEG[10] with a resolution of 2.10 Å and R-value of 0.202) obtained after minimization and equilibration processes performed using the AMBER 15 package[11] in conjunction with ff14SB protein force field.[12] in order to clean the structure and to remove bad contacts. Subsequently, the protein was solvated using the Extended Simple Point Charge (SPC/E) water model (9721 water molecules) in a truncated octahedral box with a buffer of 12 Å between the protein atoms and the edge of the box. Afterwards, a short energy minimization was performed in two steps using the Simulated Annealing with NMR-derived Energy Restraints (SANDER) module of AMBER. In the first stage, the water molecules and counter ions were relaxed with 200 cycles of minimization. In the second step, the entire system as a whole was relaxed with 1000 cycles of minimization. Subsequently, the system was heated at constant volume for 20 ps from 10 K to 300 K with 10 kcal/mol weak restraints on the protein. This process was followed by two equilibration steps, the first was performed at constant pressure (1 atm) and temperature (300 K) for 200 ps with no restraints and the second was performed in a microcanonical (NVE) ensemble for 1 ns. Relaxed scans were performed subsequently using subsystems shown in Fig. 2 where the MTSL side chain was gradually built by adding atoms to the unit peptide extracted from the crystal structure of the Aurora-A kinase.

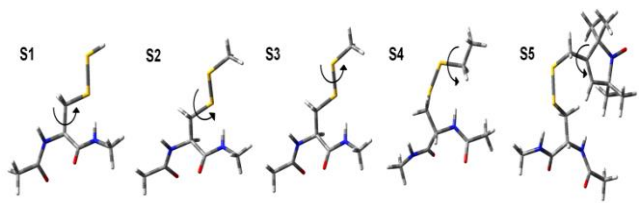

**Fig. 2:** Subsystems considered for the calculations of the torsional energy profiles about the $\chi_1$ (S1), $\chi_2$ (S2), $\chi_3$ (S3), $\chi_4$ (S4) and $\chi_5$ (S5) dihedral angles.

In order to provide reliable reproductions of the equilibrium geometries of the rotamers, the energy torsional profiles $V_i(\chi_i)$ were calculated by performing relaxed scans in thirty-seven steps of 10° with the B3LYP hybrid functional[13-14] and the 6-31G(d) basis set[15] in gas phase around each dihedral angle ($\chi_1$, $\chi_2$, $\chi_3$, $\chi_4$ and $\chi_5$) of the MTSL side chain. The same model system was used in previous work.[5-9] The *ab initio* relaxed scans were performed in 37 steps of 10 degree using the *opt = mod redundant* keyword in the Gaussian 09 software[16] that fixes coordinates but optimises or relaxes the other atoms. These scans helped to identify the minima of the torsional energy profiles of all five dihedral angles. The Gibbs free energies were determined with the B3LYP hybrid functional and the 6-31G(d) basis set using the *Freq=hindrot* keyword.

## 2. Results and discussion

### 2.1 Conformational analysis of the MTSL side chain from QM calculations

The MTSL side chain was gradually built in by adding atoms to the $C_\alpha$ atom of the $CH_3$-CO-NH-$C_\alpha$-CO-NH-$CH_3$ fragment extracted from the minimized and equilibrated structure of the Aurora-A kinase domain in which the Cartesian coordinates of the peptide atoms were kept fixed in the configuration obtained after the previous scan. The relaxed scans were performed following the strategy shown in Fig. 3. The energy minima of the $\chi_{i+1}$ dihedral angle were determined at the minima of the $\chi_i$ dihedral angle.

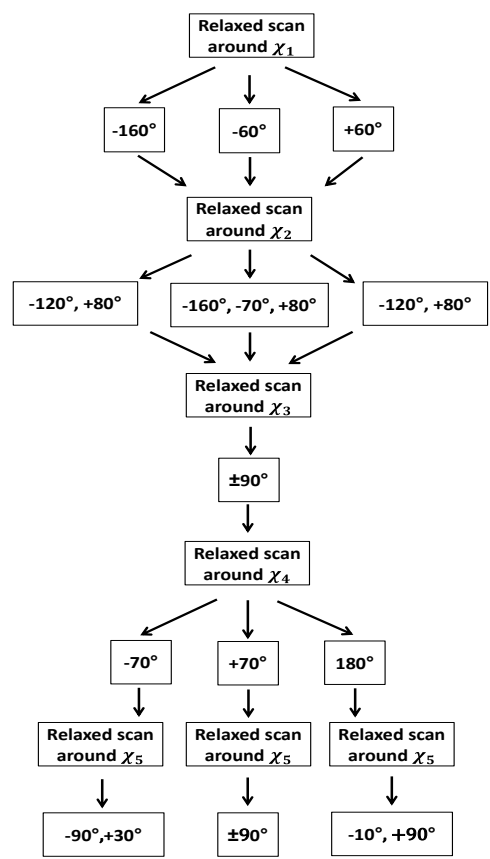

**Fig. 3:** Strategy used to find minima of the torsional energy profile around $\chi_1$, $\chi_2$, $\chi_3$, $\chi_4$ and $\chi_5$ dihedral angles from the relaxed scans. The first relaxed scan was performed around $\chi_1$ and three minima at -160°, -60° and +60° were found. Subsequently the energy torsional profile was calculated around $\chi_2$ at the minima of $\chi_1$ and two minima, one broad between -160° and -120° and another at +80° were found for $\chi_1$ = -160° and +60°. Three minima (-160°, -70° and +80°) were found for $\chi_1$ = -60°. The relaxed scan around $\chi_3$ was performed at the minima of $\chi_2$ and two minima ±90° for all the possible combination of $\chi_1$ and $\chi_2$ were found. Three minima ±70° and 180° were found in the torsional profiles of $\chi_4$. The torsional profile of $\chi_5$ showed minima depending on the values of $\chi_4$.

The first torsional energy profile of the dihedral angle, $\chi_1$ was obtained by performing relaxed scans in the range from -180° to +180°, considering the rotation of the $C_\alpha$-$C_\beta H_2$ group attached to the $CH_3$-CO-NH-$C_\alpha$H-CO-NH-$CH_3$ fragment. The torsional energy profile showed three minima at the values of -160° and ±60°(Fig. 4A).

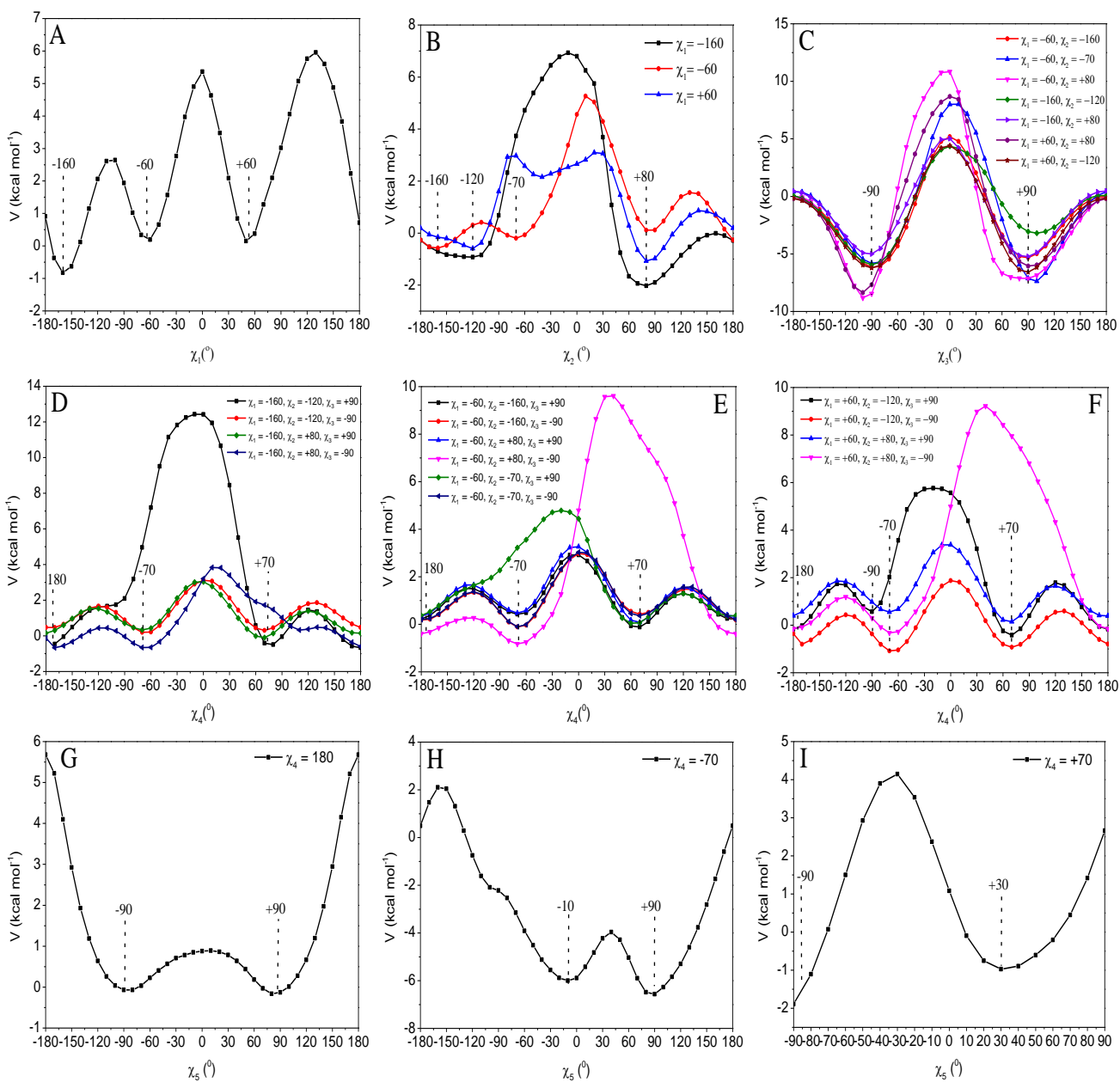

**Fig. 4:** Energy torsional profiles about the five dihedral angles, $\chi_1, \chi_2, \chi_3, \chi_4$ and $\chi_5$ of the MTSL side chain. (A) Energy torsional profile about the dihedral angle, $\chi_1$. (B) Energy torsional profiles about the dihedral angle, $\chi_2$ at the minima of $\chi_1$. (C) Energy torsional profiles about the dihedral angle, $\chi_3$ at the minima of $\chi_2$. (D-E-F) Energy torsional profile about the dihedral angle, $\chi_4$ was calculated at the minima indicated in the insert. (G) Energy torsional profile about the dihedral angle, $\chi_5$ when $\chi_4$ is equal to 180°. (H) Energy torsional profile about the dihedral angle $\chi_5$ when $\chi_4$ is equal to -70°. (I) Energy torsional profile about the dihedral angle $\chi_5$ when $\chi_4$ is equal to +70.

The high energy barriers corresponded to eclipsed configurations, while the lower energy barriers correlated to configurations in which the $S_\gamma$ and $S_\delta$ atoms were interacting with the protein backbone. The first torsional profile (Figure 4A), $V^{(1)}(\chi_1)$ showed three broad minima and two transitions between -160°↔-60° and -60°↔+60° that were separated by an energy barrier of ∼ 3 kcal/mol and ∼ 5 kcal/mol, respectively. These energy barriers are somewhat small and transitions between them would be expected to occur frequently. Subsequently, the $V^{(2)}(\chi_2)$ torsional profiles were calculated considering the rotation around the $C_\beta H_2$-$S_\gamma$ group attached to the $C_\alpha$ atom of the $CH_3$-NH-CO-$C_\alpha$H-NH-CO-$CH_3$ fragment. The $\chi_1$ dihedral angle was set at -160° and ±60° (minima in the previous scan). In the calculated $V^{(2)}(\chi_2)$ torsional profiles (Figure 4B) energy barriers of ~1-2 kcal/mol between -120°↔+80° for $\chi_1$ equal to -160° and +60°, and -160°↔-70° and -160↔+80° for $\chi_1$ equal to -60° were observed. These transitions would be expected to occur very frequently since they are separated by very small energy barriers, but are unlikely to do so considering the electrostatic interactions between the $S_\gamma$ sulphur atom and the protein backbone. The torsional profiles about $\chi_3$ were calculated at all these minima keeping the $\chi_1$ and $\chi_2$ dihedral angles fixed at the selected values shown in Figure 4C. For the calculation of the $V^{(3)}(\chi_3)$ torsional profiles, the rotation around the $C_\beta H_2 S_\gamma$-$S_\delta CH_2$ group attached to the $C_\alpha$ atom of the $CH_3$-NH-CO-$C_\alpha$H-NH-CO-$CH_3$ fragment was considered. Similar energy torsional profiles and two minima at ±90° separated by a higher energy barrier of ~ 14 kcal/mol were found for $V^{(3)}(\chi_3)$ for all seven possible combinations of the minima found for $\chi_1$ and $\chi_2$ (Figure 4C). The energy minima corresponded to the structures stabilized by short electrostatic interactions between atoms of the MTSL side chain and the unit peptide model as shown in Fig. 5.

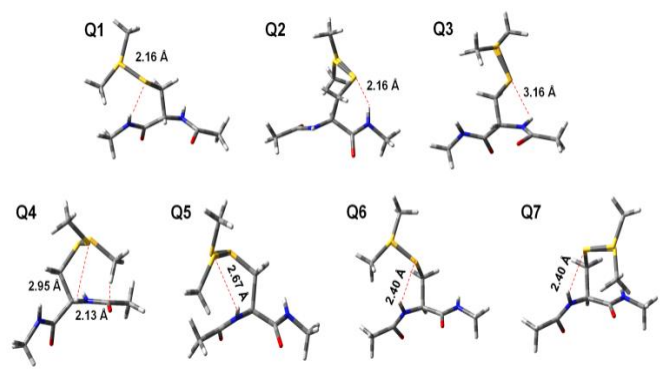

**Fig. 5:** Interactions between atoms of the MTSL side chain and the unit peptide model for different values of $\chi_1$, $\chi_2$ and $\chi_3$ (red dashed lines). Note that two conformations of $\chi_3$ are shown in each panel. (Q1) Conformation at $\chi_1$ = -160°, $\chi_2$ = -120° and $\chi_3$ = ±90°. (Q2) Conformation at $\chi_1$ = -160°, $\chi_2$ = +80° and $\chi_3$ = ±90°. (Q3) Conformation at $\chi_1$ = -60°, $\chi_2$ = -160° and $\chi_3$ = ±90°. (Q4) Conformation at $\chi_1$ = -60°, $\chi_2$ = -70° and $\chi_3$ = ±90. (Q5) Conformation at $\chi_1$ = -60°, $\chi_2$ = +80° and $\chi_3$ = ±90°. (Q6) Conformation at $\chi_1$ = +60°, $\chi_2$ = -120° and $\chi_3$ = ±90°. (Q7) Conformation at $\chi_1$ = +60°, $\chi_2$ = +80° and $\chi_3$ = ±90°.

The $S_\gamma$ atom interacts with the NH group for $\chi_1$ = -160° at $\chi_2$ = -120° (Figure 5Q1), +80° (Figure 5Q2) and for $\chi_1$ = -60° at $\chi_2$ = -160° (Figure 5Q3). The conformation at $\chi_1$ = -60°, $\chi_2$ = -70° and $\chi_3$ = +90° was stabilized by two short electrostatic interactions between the $S_\delta$ atom of the MTSL chain and the $C_\alpha$H group in the fragment, and the $CH_2$ group and the CO group (Figure 5Q4). The $S_\delta$ atom interacts with the NH group for $\chi_1$ = -60° and $\chi_2$ = +80° (Figure 5Q5). The conformation at $\chi_1$ = +60°, $\chi_2$ = -120° and +80° and $\chi_3$ = ±90° (Figures 5Q6 and Q7) was stabilized by interactions between the $S_\gamma$ atom of the MTSL and the NH group of the protein backbone. The remaining $V^{(4)}(\chi_4)$ and $V^{(5)}(\chi_5)$ torsional profiles (Figure 4D - I) were calculated considering the rotations around the $S_\delta$-$CH_2$-group (for $\chi_4$) and $S_\delta$-$CH_2$-pyrroline nitroxide ring (for $\chi_5$) with the previous dihedral angles kept fixed at the selected values. Like for $V^{(3)}(\chi_3)$, the $V^{(4)}(\chi_4)$ torsional profiles (Figures 4D, E and F) were found to be independent from the values of the previous dihedral angles, probably due to the distance from the protein backbone. Similar torsional profiles for $V^{(4)}(\chi_4)$ and three minima (±70° and 180°) were observed for all possible combinations of $\chi_1, \chi_2$ and $\chi_3$. Two low energy barriers of 1-2 kcal/mol were found between the minima at ±70° and 180° and one high energy barrier between -70° and +70° was found. The $V^{(5)}(\chi_5)$ torsional profile was measured at ±70° and 180° and three different profiles were observed (Figure 4G, H and I). The $V^{(5)}(\chi_5)$ profile at $\chi_4$=+70° was calculated between -90° to +90° due to a clash between one of the methyl groups of the pyrroline nitroxide ring and the $C_\beta$ carbon of the MTSL chain for angles over this range. The shape of the torsional profiles obtained in this work is similar to those seen in α-helices and values of the minima were found to be only slightly different,[5-9] indicating the protein backbone does not significantly influence the torsional profiles. No relevant changes in the torsional profile of $\chi_1$ and $\chi_2$ were observed on extension of the atoms in the unit peptide (data not shown) but longer computational times. This was also observed for $\chi_3$, $\chi_4$ and $\chi_5$.

After calculation of the torsional energy profiles for all five dihedral angles in the unit peptide model, a population of 76 conformations was found at the minima of the torsional energy profiles. The torsional profiles showed low energy barriers from 1 kcal/mol to ~20 kcal/mol and isoenergetic minima in the potential energy surface, indicating that the rotamer population would be fully sampled at room temperature and conformational states are expected to be obtained in the same amount. Also, the Gibbs free energy of the different conformations were seen to be comparable and equal to -6556624 kcal mol$^{-1}$, -6556624 kcal mol$^{-1}$ and -6556623 kcal mol$^{-1}$ for Q2, Q5 and Q7 (characterized by different $\chi_1$), respectively. Similarly it was observed for Q1 and Q2 (characterized by different $\chi_2$) with Gibbs free energy equal to -6556624 kcal mol$^{-1}$ and -6556622 kcal mol$^{-1}$; and for conformations with $\chi_3$ equal to +90° and +90° that have a free energy equal to -655624 kcal mol$^{-1}$ and -655623 kcal mol$^{-1}$. After having determined conformers of the MTSL side chain we tested the effect of the geometric variation of the side chain on the magnetic parameters in order to exclude the contribution of the side chain to the EPR spectrum. Previous literature showed variations of $A_{zz}$ and $g_{xx}$ components upon geometrical variations (NO bond length and the CNOC out-of-plane dihedral angle) in the proxyl radical[17] and in aromatic radical rings.[18] Hence, we tested the effect of the geometry of the MTSL side chain on the magnetic properties in order to exclude any contribution of the chain on the spectrum. Six conformations were selected at the minima of the torsional profiles (Fig. 4) with different values of the dihedral angles and the magnetic parameters were computed at DFT level. Conformers showed comparable magnetic parameters and minor changes were observed in the 94 GHz EPR spectra. This indicated that the addition of the side chain and variations of its geometry do not alter significantly the spin density and shape of the molecular orbital that remained well localized on the NO moiety, like observed in the case of the proxyl radical ring described in previous works.[17]

**Conclusions and future work**

The conformational analysis of the MTSL side chain on a short fragment of the Aurora-A kinase activation loop revealed torsional profiles comparable to those observed in fragments of α-helix studied in previous work.[5-9] This indicated that the backbone structure does not influence the torsional profiles significantly. A population of 76 conformers was found at the minima of the torsional profiles and in addition, it was observed that the $\chi_3$, $\chi_4$ and $\chi_5$ do not depend on the previous two dihedral angles, suggesting that the determined set of rotamers can be considered also for other systems. This analysis can be also used to determine some starting conformations for MD simulations of the MTSL spin-labelled Aurora-A kinase or other systems. On the basis our own experience and previous work performed using MD of MTSL spin-labelled proteins,[8-9] it was observed that transitions of $\chi_4$ and $\chi_5$ are much faster than transitions of $\chi_1$, $\chi_2$ and $\chi_3$, so the starting structures can be established fixing $\chi_1$, $\chi_2$ and $\chi_3$. It was observed that conformers have comparable potential and Gibbs free energy allowing the fixing of values $\chi_1 = \pm 60°$, -160°, $\chi_2 = +80°$ and $\chi_3 = \pm 90°$ in order to perform more advanced modelling studies.

**3. Acknowledgments**

This work was supported by a studentship from Bruker Ltd. and a Cancer Research UK grant (C24461/A12772 to R.B). The authors would like to acknowledge the use of the EPSRC UK National Service for Computational Chemistry Software (NSCCS) and its staff (Dr. Alexandra Simperler and Dr. Helen Tsui for some technical advice) at Imperial College London in carrying this work. M. G. Concilio acknowledges Dr. A. Baldansuren, Dr. N. Burton and Dr. A. Simperler for useful discussions and feedback.